\documentclass[aps,nofootinbib,superscriptaddress]{revtex4}

\usepackage{graphicx}
\usepackage{dcolumn}

\usepackage[dvips]{color}
\usepackage{amsmath}
\usepackage{amssymb}

\begin{document}

\pagenumbering{arabic}

\title{Inclusive Quasi-Elastic Neutrino Reactions}

\author{J. Nieves}
\affiliation{Departamento de F\'{\i}sica Moderna,\\
  Universidad de Granada,
  Campus de Fuentenueva, S/N,
  E-18071 Granada, Spain}
\author{J. E. Amaro}
\affiliation{Departamento de F\'{\i}sica Moderna,\\
  Universidad de Granada,
  Campus de Fuentenueva, S/N,
  E-18071 Granada, Spain}
\author{M. Valverde}
\affiliation{Departamento de F\'{\i}sica Moderna,\\
  Universidad de Granada,
  Campus de Fuentenueva, S/N,
  E-18071 Granada, Spain}
\author{E. Oset}
\affiliation{ Departamento de F\'{\i}sica Te\'{o}rica and IFIC\\
  Centro Mixto Universidad de Valencia-CSIC,
  46100 Burjassot (Valencia), Spain}

\begin{abstract}
The Quasi-Elastic (QE) contribution of the nuclear
inclusive electron scattering model developed in \cite{Nucl. Phys. A Gil}
is extended to the study of electroweak Charged Current (CC) induced
nuclear reactions, at intermediate energies of interest for future
neutrino oscillation experiments. The model accounts for
long range nuclear (RPA) correlations,
Final State Interaction (FSI) and  Coulomb corrections.
RPA correlations are shown to play a crucial role in the whole
range of neutrino energies, up to 500 MeV, studied in this work.
Predictions for inclusive muon capture for different nuclei,
and for the reactions $^{12}$C$(\nu_\mu,\mu^-)X$ and
$^{12}$C$(\nu_e,e^-)X$ near threshold are also given.
\end{abstract}

\maketitle

\section{Introduction}

Neutrino properties have been object of much interest as long as they could 
provide hints of physics beyond the standard model. A sensitive way to
study the mass of the neutrino is by means of neutrino oscillations. One of 
the various experiments devoted to this topic is the atmospheric neutrinos 
detection carried out in Kamiokande and Superkamionde,
which data have given evidence on $\nu_{\mu}\to\nu_{\tau}$ 
oscillation with $10^{-3}\lesssim \Delta m^2 \lesssim 10^{-2}$ and almost
maximal mixing angle, see \cite{Rev. of Mod. Phys. Kajita} for a review.
Once this phenomena have been firmly stablished new questions arise, 
such as the role of
three flavour oscillations and the precise determination of the values of
neutrino masses and mixing parameters \cite{EstosProc}.
For the obtention of accurate results in these new experiments it is necessary 
to keep under control the sources of systematic error.
Two of the major sources of systematic errors in the sub-GeV samples of SK
experiments are the charged and neutral-current cross sections 
\cite{Phys. Lett. B Fukuda}.
Thus, if we want to cope with the requirements of this
new experiments precise nuclear interaction models must be used.

Any model aiming at describing the interaction of neutrinos with
nuclei should be firstly tested against the existing data on the
interaction of real and virtual photons with nuclei. At intermediate
energies (nuclear excitation energies ranging from about 100 MeV to
500 or 600 MeV) three different contributions should be taken into
account: i) Quasi-Elastic (QE) processes,
         ii) pion production and two body processes
from the QE region to that beyond the $\Delta(1232)$ resonance peak,
and iii) double pion production and higher nucleon resonance degrees
of freedom induced processes. The model developed in
\cite{Nucl. Phys. A Gil} (inclusive electro--nuclear reactions) and
\cite{Nucl. Phys. A Carrasco} (inclusive photo--nuclear reactions) has been
successfully compared with data at intermediate energies and it
systematically includes the three type of contributions mentioned
above. The building blocks of this model are: i) a
gauge invariant model for the interaction of real and virtual photons
with nucleons, mesons and nucleon resonances with parameters
determined from the vacuum data, and 
ii) a microscopic treatment of nuclear  effects,
including long and short range nuclear correlations \cite{Phys. Rep. Oset},
FSI, explicit meson and $\Delta (1232)$ degrees of freedom, 
two and three nucleon absorption channels, etc. 
 Finite size effects are computed from a Local Fermi Gas
(LFG) picture of the nucleus, which is an accurate approximation to deal with
inclusive processes which explore the whole nuclear
volume \cite{Nucl. Phys. A Carrasco}.
The parameters of the model are completely fixed from previous
hadron-nucleus studies: pionic atoms, elastic and inelastic
pion-nucleus reactions, $\Lambda-$hypernuclei, etc. \cite{var_pion}.
The photon coupling constants are also determined in the
vacuum. Thus the model of \cite{Nucl. Phys. A Gil} and 
\cite{Nucl. Phys. A Carrasco} has no free parameters, and hence these results
are predictions deduced from the nuclear microscopic framework developed
in \cite{Phys. Rep. Oset} and \cite{var_pion}. 
In this talk, we show an extension of the nuclear inclusive QE
electron scattering model of \cite{Nucl. Phys. A Gil},
including the axial CC current, to describe neutrino and
antineutrino induced nuclear reactions in the QE region. We will not show
here many details of the model, for a detailed discussion 
we refer the reader to \cite{Phys. Rev. C Nieves}.

\section{Inclusive cross section} \label{sec:xsec}

We will present here the general formalism focusing on the neutrino 
Charged-Current (CC) reaction
\begin{equation}
\nu_l (k) +\, A_Z \to l^- (k^\prime) + X 
\label{eq:reac}
\end{equation}
though the generalization of the obtained expressions to antineutrino
induced reactions or  inclusive muon capture in nuclei is straightforward.

The double differential cross section, with respect to the outgoing
lepton kinematical variables,  for the process of Eq.~(\ref{eq:reac})
is given in the Laboratory (LAB) frame by
\begin{equation}
\frac{d^2\sigma_{\nu l}}{d\Omega(\hat{k^\prime})dE^\prime_l} =
\frac{|\vec{k}^\prime|}{|\vec{k}~|}\frac{G^2}{4\pi^2} 
L_{\mu\sigma}W^{\mu\sigma} \label{eq:sec}
\end{equation}
with $L$ and $W$ the leptonic and hadronic tensors,
respectively. 

The hadronic tensor $W^{\mu\nu}$ includes non-leptonic
vertices and corresponds to the charged electroweak transitions of the
target nucleus to all possible final states; it is
completely determined by six independent, real Lorentz scalar 
structure functions $W_i(q^2)$ , $i=1,\ldots,6$.

\begin{figure}
 \begin{center}
  \includegraphics[scale=0.5]{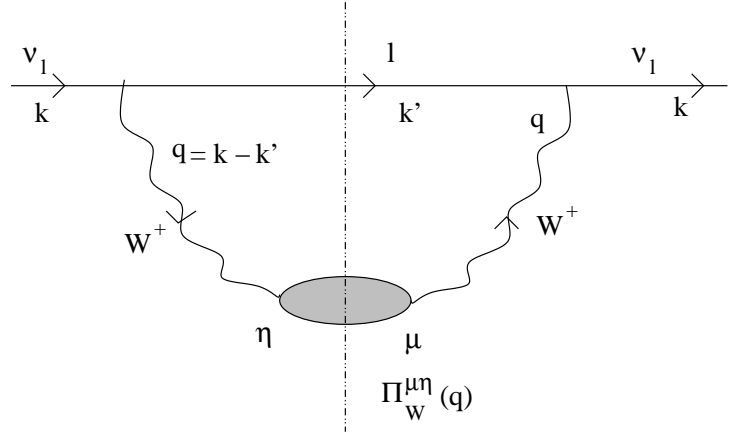}
 \end{center}
  \caption{Diagrammatic representation of the neutrino
    self-energy in nuclear matter.}\label{fig:selfen}
\end{figure}

 We follow here the formalism of~\cite{Nucl. Phys. A Gil}, and 
evaluate the self-energy $\Sigma_\nu^r(k;\rho)$ of a neutrino with
helicity $r$ inside of a nuclear medium of density $\rho$.
Diagrammatically this is depicted in Fig.\ ~\ref{fig:selfen} 
After summing over helicities, we get
\begin{equation}
\Sigma_\nu(k;\rho) = \frac{8{\rm i} G}{\sqrt 2 M^2_W} \int
\frac{d^4q}{(2\pi)^4} \frac{L_{\eta\mu} \Pi^{\mu\eta}_W(q;\rho) 
}{k^{\prime 2}-m^2_l+ {\rm i}\epsilon}
\end{equation}
where  $\Pi^{\mu\rho}_W(q)$ is the $W^+-$boson self-energy in the nuclear
medium.

The neutrino disappears from the elastic flux, by inducing 
one~particle~-~one~hole (1p1h), 2p2h $\ldots$ excitations, $\Delta(1232)-$hole
($\Delta$h) excitations, or creating pions, etc\ldots at a rate given by
\begin{equation}
\Gamma (k;\rho) = - \frac{1}{k^0 }{\rm Im} \Sigma_\nu (k;\rho)
\end{equation}
To evaluate the imaginary part of $\Sigma_\nu$ we use the Cutkosky's
rules, and we cut with a straight vertical line (see
Fig.\ \ref{fig:selfen}) the intermediate lepton state and those produced
by the $W-$boson polarization (shaded region). Those states are then
placed on shell by taking the imaginary part of the propagator,
self-energy, etc. Thus, we obtain 
\begin{equation}
{\rm Im} \Sigma_\nu(k) = \frac{8G}{\sqrt 2 M^2_W}\int \frac{d^3
  k^\prime}{(2\pi)^3 }\frac{\Theta(q^0) }{2E^{\prime}_l} 
 {\rm Im}\left\{ \Pi^{\mu\eta}_W(q;\rho) L_{\eta\mu} \right\} 
\label{eq:ims}
\end{equation}
for $k^0 > 0$.

\begin{figure}
\begin{center}\includegraphics[scale=0.3]{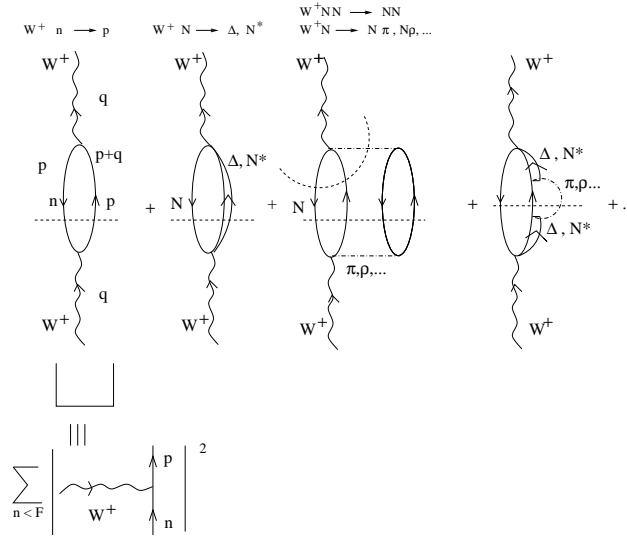}
\end{center}
\caption{ Diagrams of some processes contributing to the $W^+$ self-energy.}
\label{fig:fig2}
\end{figure}

Since $\Gamma dt dS$ provides a probability times a differential of area,
which is a contribution to  the $(\nu_l,l^-)$ cross section, we have
\begin{eqnarray}
d\sigma &=& \Gamma(k;\rho)dtdS = - \frac{1}{k^0 }{\rm Im} \Sigma_\nu
(k;\rho) dtdS = - \frac{1}{|\vec{k}|} {\rm Im} \Sigma_\nu (k;\rho)
d^3r
\end{eqnarray}
so the nuclear cross section is given by
\begin{equation}\label{eq:sigLDA}
\sigma = - \frac{1}{|\vec{k}|} \int {\rm Im} \Sigma_\nu (k;\rho(r)) d^3r
\end{equation}
where we are considering $\Sigma_\nu$ a function of the nuclear
density $\rho(r)$ at each point of the nucleus and we integrate over the whole
nuclear volume. We assume LDA, which, as shown 
in \cite{Nucl. Phys. A Carrasco}, is an excellent approximation
for volume processes like the one studied here. 
Coming back to Eq.\ (\ref{eq:sigLDA}) we can compare it with
Eq.\ (\ref{eq:sec}) so the hadronic tensor 
($W^{\mu\sigma}=W^{\mu\sigma}_s+iW^{\mu\sigma}_a$) reads
\begin{eqnarray}
W^{\mu\sigma}_s &=& - \Theta(q^0) \left (\frac{2\sqrt 2}{g} \right )^2 
\int \frac{d^3 r}{2\pi}~ {\rm Im}\left [ \Pi_W^{\mu\sigma} 
+ \Pi_W^{\sigma\mu} \right ] (q;\rho)\label{eq:wmunus}\\
W^{\mu\sigma}_a &=& - \Theta(q^0) \left (\frac{2\sqrt 2}{g} \right )^2
 \int \frac{d^3 r}{2\pi}~{\rm Re}\left [ \Pi_W^{\mu\sigma} 
- \Pi_W^{\sigma\mu}\right] (q;\rho) \label{eq:wmunua}
\end{eqnarray}
from where we can see how $\Pi_W^{\mu\sigma} $ is the basic object
of our approach. Following the lines of
\cite{Nucl. Phys. A Gil}, we should perform a many body expansion,
where the relevant gauge-boson absorption modes would be systematically
incorporated: absorption by one, two or even
three nucleon mechanisms, real and virtual meson ($\pi$, $\rho$,
$\cdots$) production, excitation of $\Delta$ or higher resonance
degrees of freedom, etc. Some of these modes are depicted in
Fig.\ \ref{fig:fig2} 

\section{QE contribution to $\Pi_W^{\mu\sigma}$}\label{sec:self}

The virtual $W^+$ can be absorbed by one nucleon leading to the QE
contribution of the nuclear response function. Such a contribution
corresponds to a 1p1h nuclear excitation (first of the diagrams
depicted in Fig.\ \ref{fig:fig2}). 
We will work on a non-symmetric nuclear matter with
different Fermi sea levels for protons than for neutrons.
For the $W^+\mbox{-}pn$ vertex we consider the $V-A$ current, and use
PCAC and invariance under G-parity  to relate the
pseudoscalar form factor to the axial one and to discard a term of the
form $(p^\mu+p^{\prime \mu})\gamma_5$ in the axial sector,
respectively.  Invariance under time reversal guarantees that all form
factors are real. Using isospin symmetry we can relate the vector
form factors with the electromagnetic ones. 

With all of these ingredients is straightforward to evaluate the 
contribution to the $W^+$ self-energy  of the
first  diagram of Fig.\ \ref{fig:fig2} 
We finally get 
\begin{equation}\label{eq:res}
\begin{split}
W^{\mu\nu}(q^0,\vec{q}\,)   &= - \frac{\cos^2\theta_C}{2M^2}
 \int_0^\infty dr r^2  2 \Theta(q^0) 
 \int \frac{d^3p}{(2\pi)^3}\frac{M}{E_{\vec{p}}}
      \frac{M}{E_{\vec{p}+\vec{q}}} (-\pi)  \\
 & \quad\times  \Theta(k_F^n-|\vec{p}~|) \Theta(|\vec{p}+\vec{q}~|-k_F^p)   
   \delta(q^0 + E_{\vec{p}} - E_{\vec{p}+\vec{q}}) 
  A^{\nu\mu}(p,q)|_{p^0=E_{\vec{p}}}   \\
\end{split}
\end{equation}
with the CC nucleon tensor $A^{\mu\nu}$ obtained after taking
some traces on the Dirac's space.
The $d^3p$ integrations above can be done analytically and all of them
are determined by the imaginary part of the relativistic isospin
asymmetric Lindhard function, $\overline {U}_R(q,k_F^n,k_F^p)$.  
Explicit expressions for $\overline{U}_R$ and $A^{\mu\nu}$ are given 
in \cite{Phys. Rev. C Nieves}.

Up to this point the treatment is fully relativistic. 
To account for RPA effects, we will
use a nucleon--nucleon effective force, so for consistency we ought to use a
non-relativistic Fermi gas. This is easily done by replacing the 
factors $M/E_{\vec{p}}$ and $M/E_{\vec{p}+\vec{q}}$ in
Eq.~(\ref{eq:res}) by one.

Pauli blocking, through the imaginary part of the Lindhard function, is
the main nuclear effect included in the hadronic tensor of
Eq.~(\ref{eq:res}). In the next sections we will study additional
nuclear corrections to $W^{\mu\nu}$.

A few words here on the low density theorem (LDT): when low nuclear density 
is supposed, the imaginary part of the Lindhard function can be approximated
by a Dirac delta on  energy (up to a constant factor) in such a way that the
model reproduces the free space nucleon cross section.

\section{Nuclear Model Corrections}

 \subsection{Nuclear Correlations}
\begin{figure}
\begin{center}\includegraphics[scale=0.3]{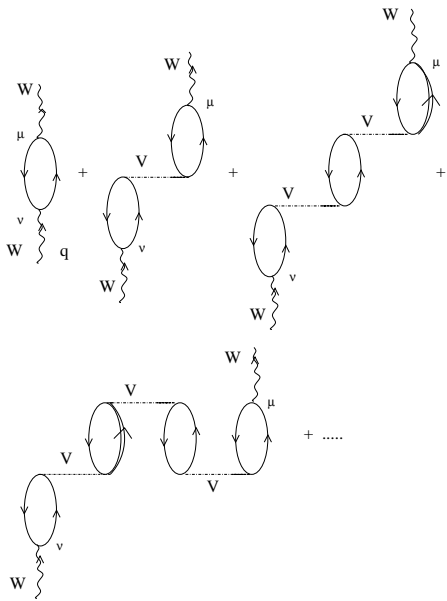}
\end{center}
\caption{Set of irreducible diagrams responsible for the
  polarization (RPA) effects in the 1p1h contribution to the
  $W^+$ self-energy. }\label{fig:fig3}
\end{figure}

   When the electroweak interactions take place in nuclei the strengths
of electroweak couplings may change from their free nucleon values due
to the presence of strongly interacting nucleons; indeed,
since the nuclear experiments on $\beta$ decay in the early seventies
\cite{Nucl. Phys. A Wilkinson}, the quenching of axial current is a well
established phenomenon. We follow here the many body approach of 
\cite{Nucl. Phys. A Gil}, and take into account the medium polarization 
effects in the 1p1h contribution to the $W^+$ boson self-energy
by substituting it with an RPA response as shown diagrammatically in
Fig.\ \ref{fig:fig3} For that purpose we use an effective ph--ph
contact interaction 
\begin{equation}
\begin{array}{ll}
V = & c_{0}\left\{
f_{0}(\rho)+f_{0}^{\prime}(\rho)\vec{\tau}_{1}\vec{\tau}_{2}+ 
g_{0}(\rho)\vec{\sigma}_{1}\vec{\sigma}_{2}+g_{0}^{\prime}(\rho)
\vec{\sigma}_{1}\vec{\sigma}_{2}
\vec{\tau}_{1}\vec{\tau}_{2}
\right\}
\end{array}
\end{equation}
of the Landau-Migdal type.
The density dependent coefficients were determined \cite{Speth77} 
from calculations
of nuclear electric and magnetic moments, transition probabilities, and giant
electric and magnetic multipole resonances. In the $S = T = 1$ channel
($\vec{\sigma} \vec{\sigma} \vec{\tau} \vec{\tau}$ operator) we use an
interaction with explicit $\pi$ (longitudinal) and $\rho$ (transverse)
exchanges, which has been used for the renormalization of the pionic
and pion related channels in different nuclear reactions at
intermediate energies \cite{Nucl. Phys. A Gil,Nucl. Phys. A Carrasco}.
Further effects such as short range correlations (SRC)
are also taken into account.

We also include $\Delta(1232)$ degrees of freedom in the nuclear
medium which, given the spin-isospin quantum numbers of the $\Delta$
resonance, only modify the vector-isovector ($S = T = 1$) channel of
the RPA response function.

The $V$ lines in Fig.\ \ref{fig:fig3} stand for the
effective ph($\Delta$h)-ph($\Delta$h)  interaction described so
far. We should stress that this effective interaction is non-relativistic, 
and then for consistency we will  neglect terms of order
${\cal O}(p^2/M^2)$ when summing up the RPA series. 

 \subsection{Energy Balance and Coulomb Distortion}

  To ensure the correct energy balance in the reaction~(\ref{eq:reac})
for finite nuclei, the energy conserving Dirac delta function in
Eq.\ (\ref{eq:res}) has to be modified by including the minimum excitation
energy, $Q=M(A_{Z+1})-M(A_Z)$, needed for the transition to the ground state
of the final nucleus. The consideration of this energy gap is essential to
obtain reasonable cross sections for low-energy neutrinos, 
see~\cite{Oset_neut}.

We also include a Coulomb self-energy $\Sigma_C = 2k^{\prime 0} V_C(r)$ 
in the intermediate lepton propagator of the neutrino self-energy depicted in
Fig.\ \ref{fig:selfen} where $V_C(r)$ is the nucleus Coulomb potential 
produced by a charge distribution $\rho_{ch}(r)$.
This way of taking into account the
Coulomb effects has clear resemblances with what is called ``modified
effective momentum approximation'' in \cite{En98}.

 \subsection{FSI}

Once a ph excitation is produced by the virtual $W-$boson, the
outgoing nucleon can collide many times, thus inducing the emission of
other nucleons. The result is a quenching of the
QE peak respect to the simple ph excitation calculation and a
spreading of the strength, or widening of the peak.
A distorted wave approximation with an optical (complex)
nucleon-nucleus potential would remove all these events. However, if
we want to evaluate the inclusive $(\nu_l,l^-)$ cross section these
events should be kept and one must sum over all open final state
channels. 

\begin{figure}
\begin{center}
\includegraphics[scale=0.8]{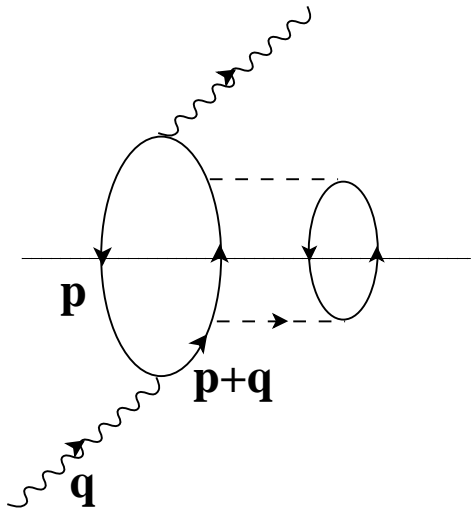}
\end{center}
\caption{$W^+$ self-energy diagram obtained from the
  first diagram depicted in Fig.\ \protect\ref{fig:fig2} by dressing up
  the nucleon propagator of the particle state in the ph
  excitation. }\label{fig:burbuja}
\end{figure}

We will account for the Final State Interaction (FSI) by using
nucleon propagators properly dressed with a realistic self-energy in
the medium, which depends explicitly on the energy and the
momentum~\cite{Phys. Rev. C Fdez}. 
This self-energy has an imaginary part from the coupling to the 2p2h 
components, which is
equivalent to the use of correlated wave functions, evaluated from
realistic $NN$ forces and incorporating the effects of the nucleon
force in the nucleon pairs. Thus, we consider the many body diagram
depicted in Fig.\ \ref{fig:burbuja}

Once we have got a model for the nucleon self-energy $\Sigma (p^0 , \vec{p}\,;
 \rho)$, we can include in it a renormalized nucleon propagator 
$G_{\rm FSI}(p;\rho)$,
that can be easily related to $S_p$ and $S_h$, the particle and
hole spectral functions and then through 
\begin{equation}
{\rm Im} \overline{U}_{FSI}(q;k_F) = 
-\frac{\Theta(q^0)}{4\pi^2}\int d^3p 
 \int_{\mu-q^0}^\mu d\omega S_h(\omega,\vec{p}\,;\rho)
S_p(q^0+\omega,\vec{p}+\vec{q}\,;\rho) 
\end{equation}
with the Lindhard function that we include in our formalism.

\section{Low Energy Results}

\begin{figure}
  \begin{center}
    \includegraphics[scale=0.3]{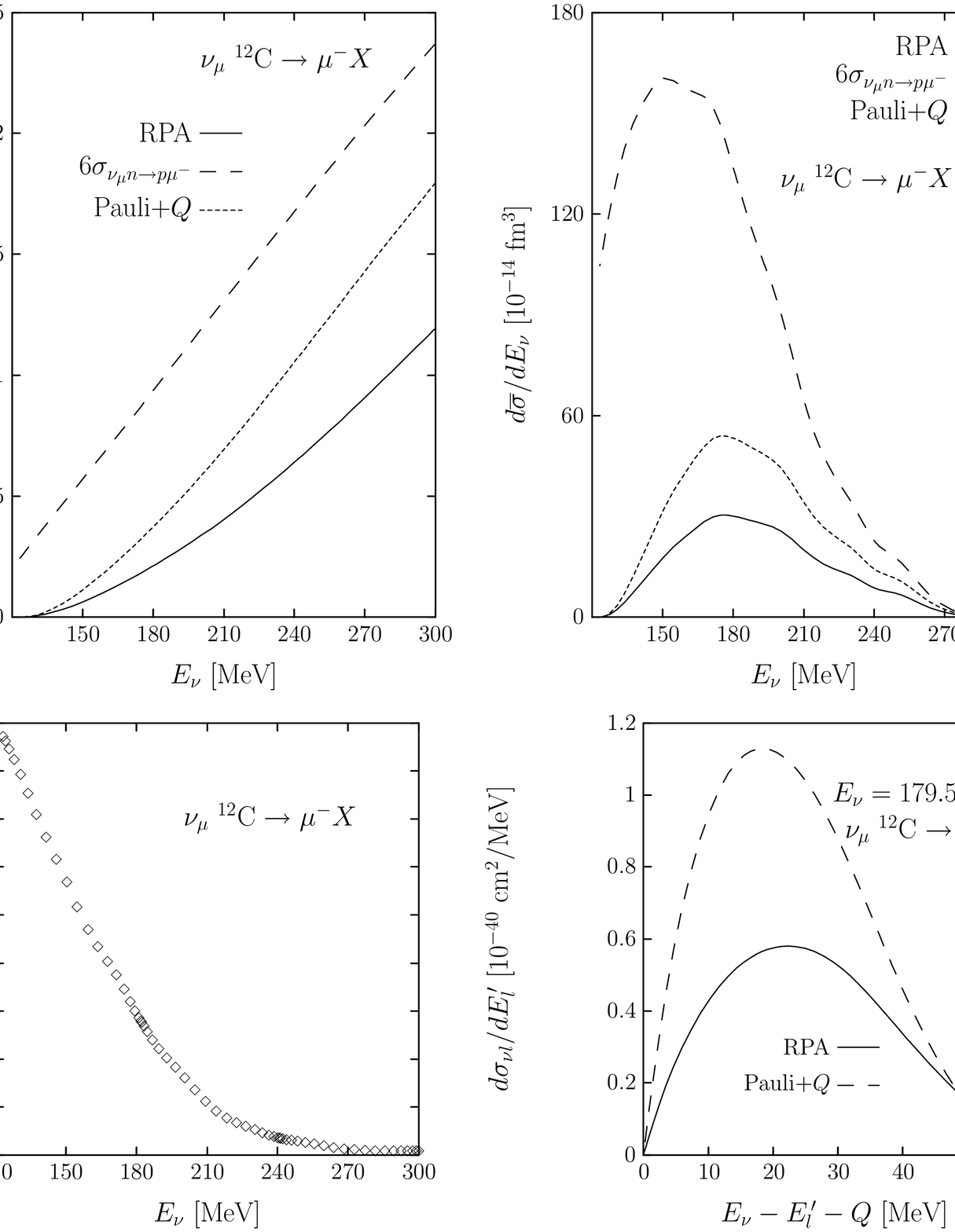}
  \end{center}
  \caption{Predictions for the LSND experiment. See \cite{Phys. Rev. C Nieves} for details.}
  \label{fig:mubaja}
\end{figure}

\begin{figure}
   \begin{center}
\includegraphics[scale=0.3]{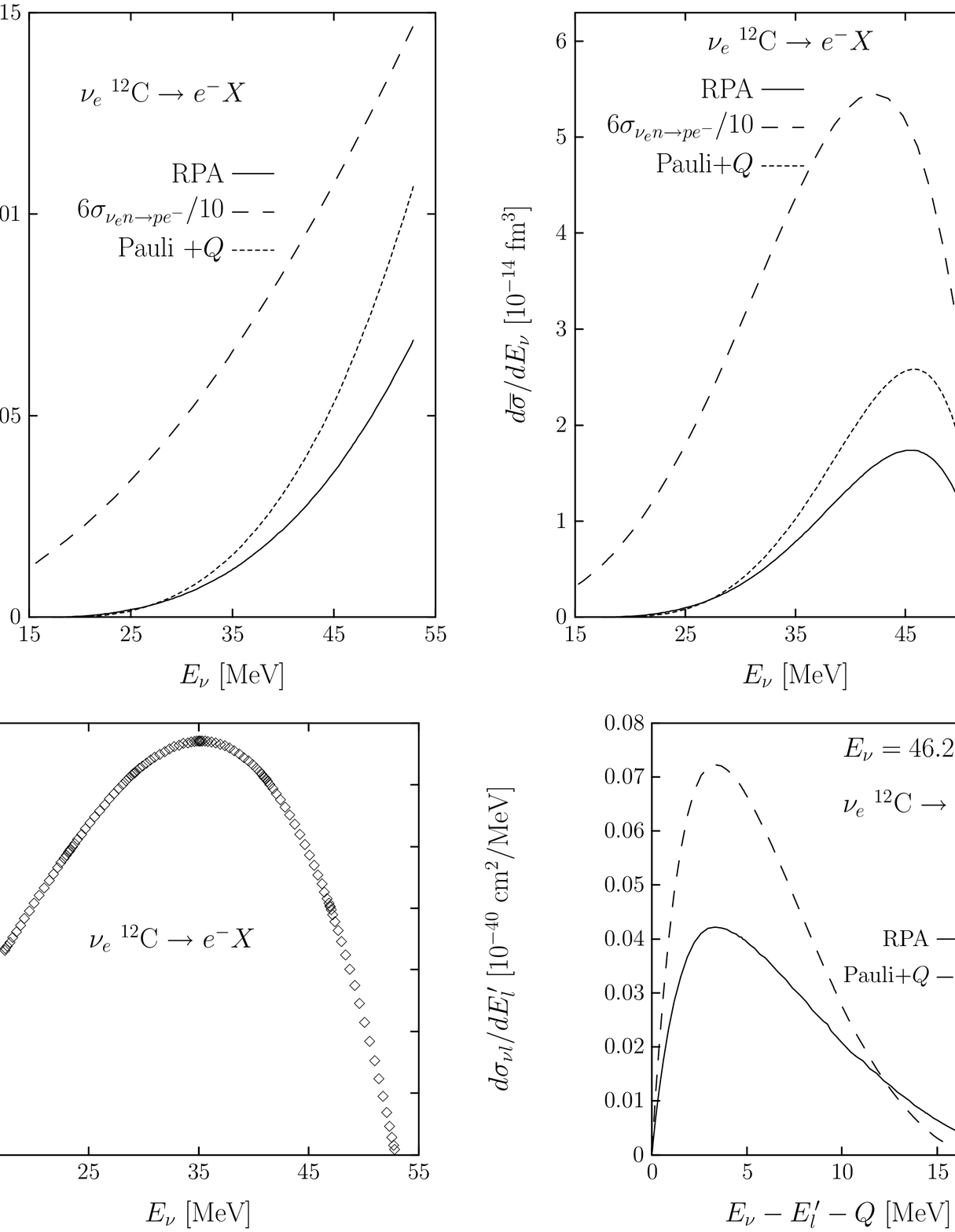}
   \end{center}
   \caption{Predictions for the $^{12}{\rm C}(\nu_e,e^-)X$ reaction. See \cite{Phys. Rev. C Nieves} for details.}
   \label{fig:elecbaja}
\end{figure}

We present in Figs.\ \ref{fig:mubaja} and \ref{fig:elecbaja}
and Table~\ref{tab:lsnd} our theoretical predictions and a comparison with
the experimental measurements of the inclusive $^{12}$C$(\nu_\mu,\mu^-)X$ 
and $^{12}$C$(\nu_e,e^-)X$ reactions near
threshold.  Pauli blocking and the use of the correct
energy balance improve the results, but only once RPA and Coulomb
effects are included a good description of data is achieved.

\begin{table}
 \caption{Flux averaged
 $^{12}{\rm C}(\nu_e,e^-)X$ and $^{12}{\rm C}(\nu_\mu,\mu^-)X$ cross sections.}
 \label{tab:lsnd}
\begin{tabular}{lcccc}\hline
& Theory & KARMEN~\cite{PLB_KAR} & LSND~\cite{PRC_LSND} 
& LAMPF~\cite{PRC_LAMPF}     \\ \hline
$\overline{\sigma}_e$  &  ~0.14~ & ~$0.15\pm 0.01 \pm 0.01$~ & ~$0.15 \pm 0.01 \pm 0.01$ ~& ~$0.141 \pm 0.023$~                          \\ \hline 
& Theory  & LSND'95 & LSND'97 & LSND'02 ~\cite{PRC_LSND-II} \\ \hline
$\overline{\sigma}_{\mu}$  &  ~11.9~ & ~$8.3\pm 0.7 \pm 1.6$~ &
 ~$11.2 \pm 0.3 \pm 1.8$ ~& ~$10.6 \pm 0.3 \pm 1.8 $~ \\\hline 
\end{tabular}
\end{table}

Given the succes of the model at low energies we decided to further test it by
calculating inclusive muon capture rates in nuclei throughout 
the Periodic Table. 
Results are given in Table~\ref{tab:capres} including the error in the 
theoretical predictions. Data were taken from
\cite{Suzuki87}, using a weighted average:
$\overline{\Gamma}/\sigma^2 = \sum_i \Gamma_i/\sigma_i^2$, with
$1/\sigma^2 = \sum_i 1/\sigma_i^2$.

\begin{table}
\caption{Experimental and theoretical total muon capture widths $\Gamma$,
 for different nuclei. See \cite{Phys. Rev. C Nieves} for details.}
  \label{tab:capres}
  \begin{center}
   \begin{tabular}{c|cccc}\hline
& Pauli+$\overline{Q}$ & RPA & Exp & $\left(\Gamma^{\rm Exp} -\Gamma^{\rm Th}
\right )/\Gamma^{\rm Exp} $ \\\hline  $^{12}$C & 5.42 & 3.21 &
$3.78\pm 0.03$ & \phantom{$-$}0.15 \\ $^{16}$O & 17.56 & 10.41 &
$10.24\pm 0.06$ & $-0.02$ \\ $^{18}$O & 11.94 & 7.77 & $8.80\pm 0.15$
& \phantom{$-$}0.12 \\ $^{23}$Na &58.38 & 35.03 & $37.73\pm 0.14$ &
\phantom{$-$}0.07 \\ $^{40}$Ca &465.5 &257.9 &$252.5\pm 0.6 $ &
$-0.02$ \\ $^{44}$Ca &318 &189 & $179 \pm 4 $ & $-0.06$ \\ $^{75}$As
&1148 & 679 & 609$\pm 4$ & $-0.11$ \\ $^{112}$Cd &1825 & 1078 &
1061$\pm 9 $ & $-0.02$ \\ $^{208}$Pb & 1939 & 1310 & 1311$\pm 8 $ &
\phantom{$-$}$0.00$ \\ \hline 
  \end{tabular}
 \end{center} 
\end{table} 

Despite the huge range of variation of the capture widths,
the agreement to data is quite good for all studied nuclei,
with discrepancies of about 15\% at most.
Furthermore, using LFG instead of a more refined model such as a shell model
does not affect much the value of integrated observables such as total 
cross section or capture widths, see \cite{Chiara05}.
It is precisely for $^{12}$C where we find the greatest discrepancy with
experiment. Nevertheless, our model provides one of the best existing
combined description of the inclusive muon capture in $^{12}$C and the
LSND measurement of the reaction $^{12}$C$(\nu_\mu,\mu^-)X$ near threshold. 

\section{Intermediate Energy Results}

At intermediate energies the predictions of this model should
become reliable, not only for integrated,
but also for differential cross sections. 
We present results for incoming neutrino energies within the
interval 150-400 (250-500) MeV for electron (muon) species. The use of
relativistic kinematics for the nucleons leads to moderate reductions in the
interval of 4-9\% for both neutrino and antineutrino cross sections, 
at the energies considered.
Such corrections do not depend significantly on the considered nucleus.

\begin{figure}
\begin{center}
\includegraphics[scale=0.6]{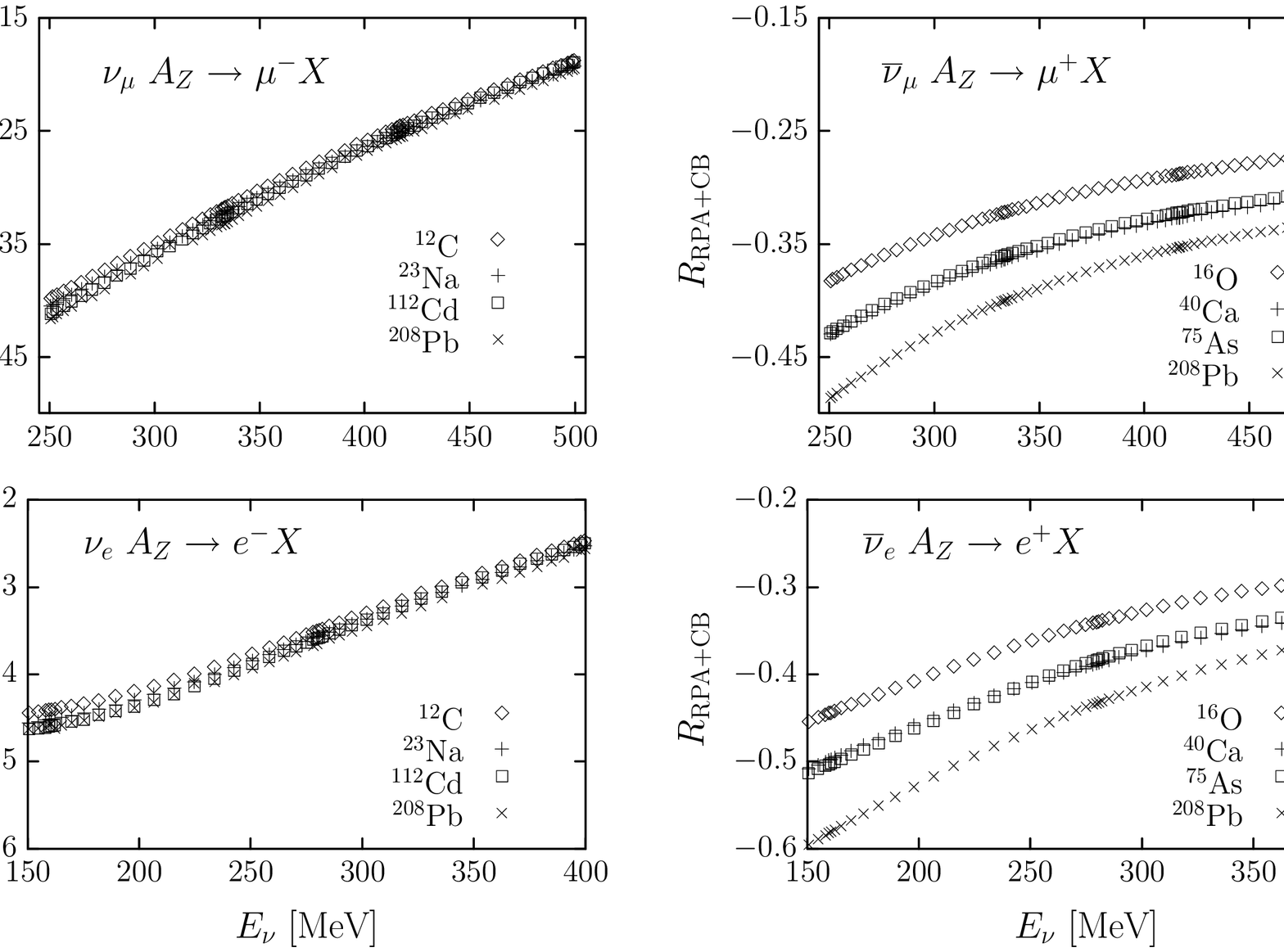}
\end{center}
\vspace{-7cm}
\caption{RPA and Coulomb (CB) corrections to electron
  and muon neutrino and antineutrino QE cross sections for different
  nuclei, as a function of the neutrino energy.}
  \label{fig:corrections}
\end{figure}

In Fig.\ \ref{fig:corrections} the effects of RPA and Coulomb
corrections are studied as a function of the incoming
neutrino/antineutrino energy. 
The correction $R_{\rm RPA+CB}$ is defined as 
$(\sigma_{\rm RPA+CB}-\sigma_0)/\sigma_0$, where
$\sigma_{0}$ does not include RPA  and Coulomb corrections,
while $\sigma_{\rm RPA+CB}$  includes these nuclear effects.
FSI corrections are not taken into account in these cross sections.
RPA correlations reduce the cross
sections, and we see large effects, specially at lower energies.  
Nevertheless, for the
highest energies considered (500 and 400 MeV for muon and electron
neutrino reactions, respectively) we still find suppressions of about
20-30\%.  Coulomb distortion of the outgoing charged lepton enhances
(reduces) the cross sections for neutrino (antineutrino) processes and its
effects decrease with energy. For antineutrino reactions,
the combined effect of RPA and Coulomb corrections have a moderated 
dependence on $A$ and $Z$. 
At the high energy end the $A-$dependence becomes milder, 
since Coulomb distortion becomes less important. 
In the case of neutrinos, the increase of the
cross section due to Coulomb cancels out partially with the RPA
reduction. Finally, the existing differences between electron and muon
neutrino/antineutrino plots are due to the different momenta of an
electron and a muon with the same energy.

\begin{figure}
\begin{center}
\includegraphics[scale=0.5]{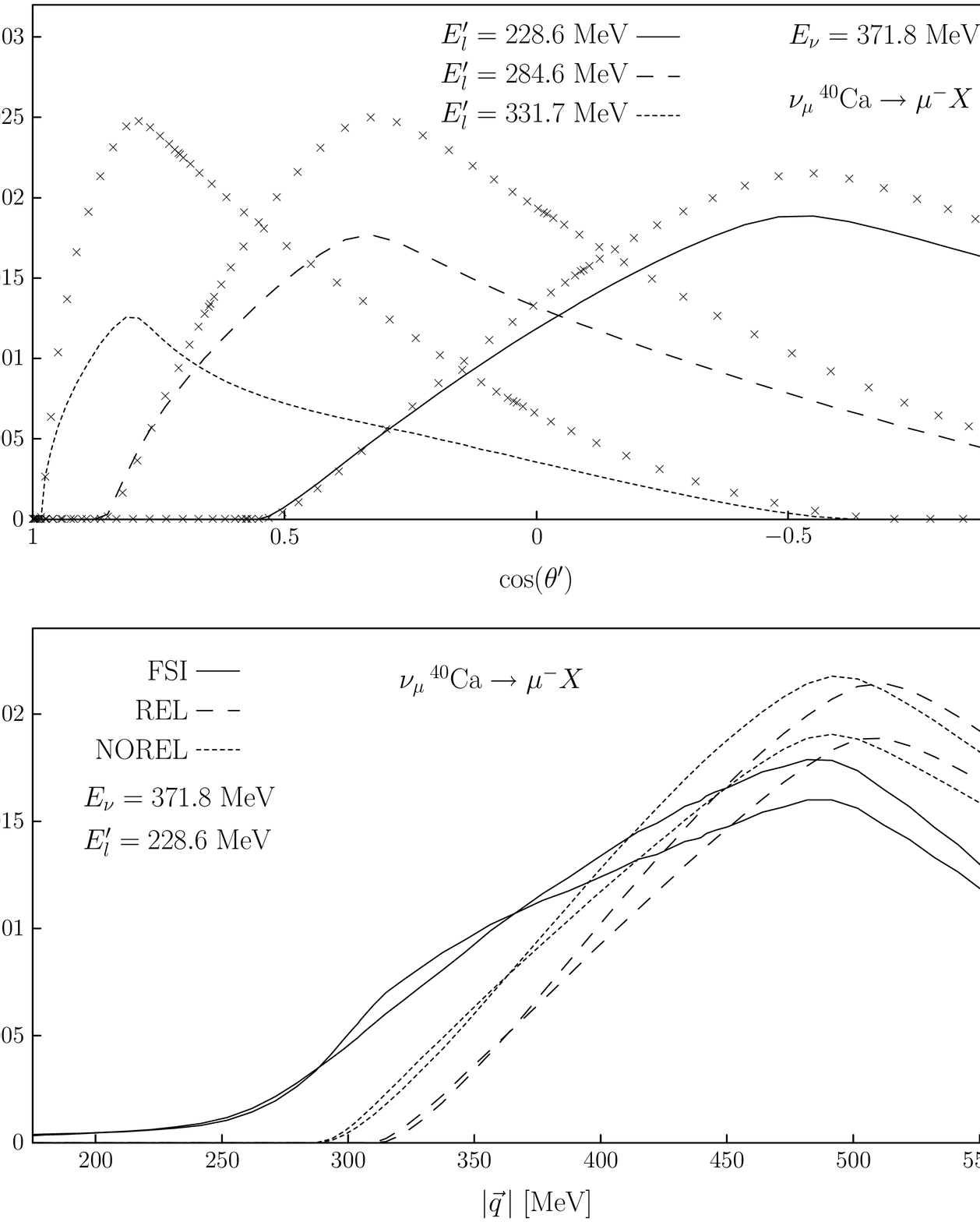}
\end{center}
\vspace{-1cm}
\caption{Muon neutrino differential cross sections in
  calcium as a function of the lepton scattering angle (top) and of
  the momentum transfer (bottom). 
  {\bf Top}: Cross sections, without FSI and using relativistic
  kinematics for the nucleons. Crosses
  have been obtained without RPA and Coulomb effects, while the curves
  have been obtained with the full model (up to FSI effects).
  {\bf Bottom}: Cross sections, obtained by
  using relativistic (REL) and
  non-relativistic nucleon kinematics results 
  with (FSI) and without (NOREL)
  FSI effects. We also  take into account RPA
  and Coulomb corrections (lower lines at the peak).}\label{fig:secdouble}
\end{figure}

\begin{table}
\caption{Muon and electron neutrino and
  antineutrino inclusive QE  integrated cross sections from $^{16}$O.}
\label{tab:fsi}
 \begin{center} 
\begin{tabular}{cr|ccc|ccc}\hline
$E_\nu$ & &\multicolumn{3}{c|}{$\sigma\left(^{16}{\rm O}(\nu_\mu,\mu^-)X
\right)$~}&\multicolumn{3}{c}{$\sigma \left ( ^{16}{\rm O} ({\bar
      \nu}_\mu,\mu^+)X\right )$~} \\
\hline
        &         & ~~REL  &~~NOREL  & FSI  &~~REL   &~~NOREL  & FSI  \\\hline
 500    & Pauli ~ &~~460.0 &~~497.0  &431.6 &~~155.8 &~~168.4  &149.9 \\ 
        & RPA   ~ &~~375.5 &~~413.0  &389.8 &~~113.4 &~~126.8  &129.7 \\\hline
 375    & Pauli ~ &~~334.6 &~~354.8  &292.2 &~~115.1 &~~122.6  &105.0 \\
        & RPA   ~ &~~243.1 &~~263.9  &243.9 &~~79.8  &~~87.9   &87.5  \\\hline
 250    & Pauli ~ &~~155.7 &~~162.2  &122.5 &~~63.4  &~~66.4   &52.8  \\  
        & RPA   ~ &~~ 94.9 &~~101.9  &93.6  &~~38.8  &~~42.1   &40.3  \\\hline
\\\hline
$E_\nu$ & &\multicolumn{3}{c|}{$\sigma\left(^{16}{\rm O}(\nu_e,e^-)X\right)$~} 
&\multicolumn{3}{c}{$\sigma \left ( ^{16}{\rm O}({\bar
      \nu}_e,e^+)X\right )$~} \\
\hline
        &         &~~ REL    & ~~NOREL & FSI  & ~~REL & ~~NOREL & FSI  \\\hline
 310    & Pauli ~ & ~~281.4  & ~~297.4 &240.6 &~~98.1 &~~ 104.0 & 87.2 \\
        & RPA   ~ & ~~192.2  & ~~209.0 &195.2 &~~65.9 & ~~72.4  & 73.0 \\\hline
 220    & Pauli ~ & ~~149.5  & ~~156.2 &121.2 &~~60.7 & ~~63.6  & 51.0 \\
        & RPA   ~ & ~~ 90.1  & ~~ 97.3  &92.8  &~~36.8 & ~~40.0 & 40.2 \\\hline
 130    & Pauli ~ & ~~37.0   & ~~ 38.3  &28.8  &~~21.1 &~~21.9  & 16.9 \\
        & RPA   ~ & ~~20.6   & ~~22.3  & 23.3 &~~10.9 & ~~11.9  & 12.8 \\\hline
\end{tabular}
\end{center} 
\end{table} 

\begin{figure}
\begin{center}
\includegraphics[scale=0.45]{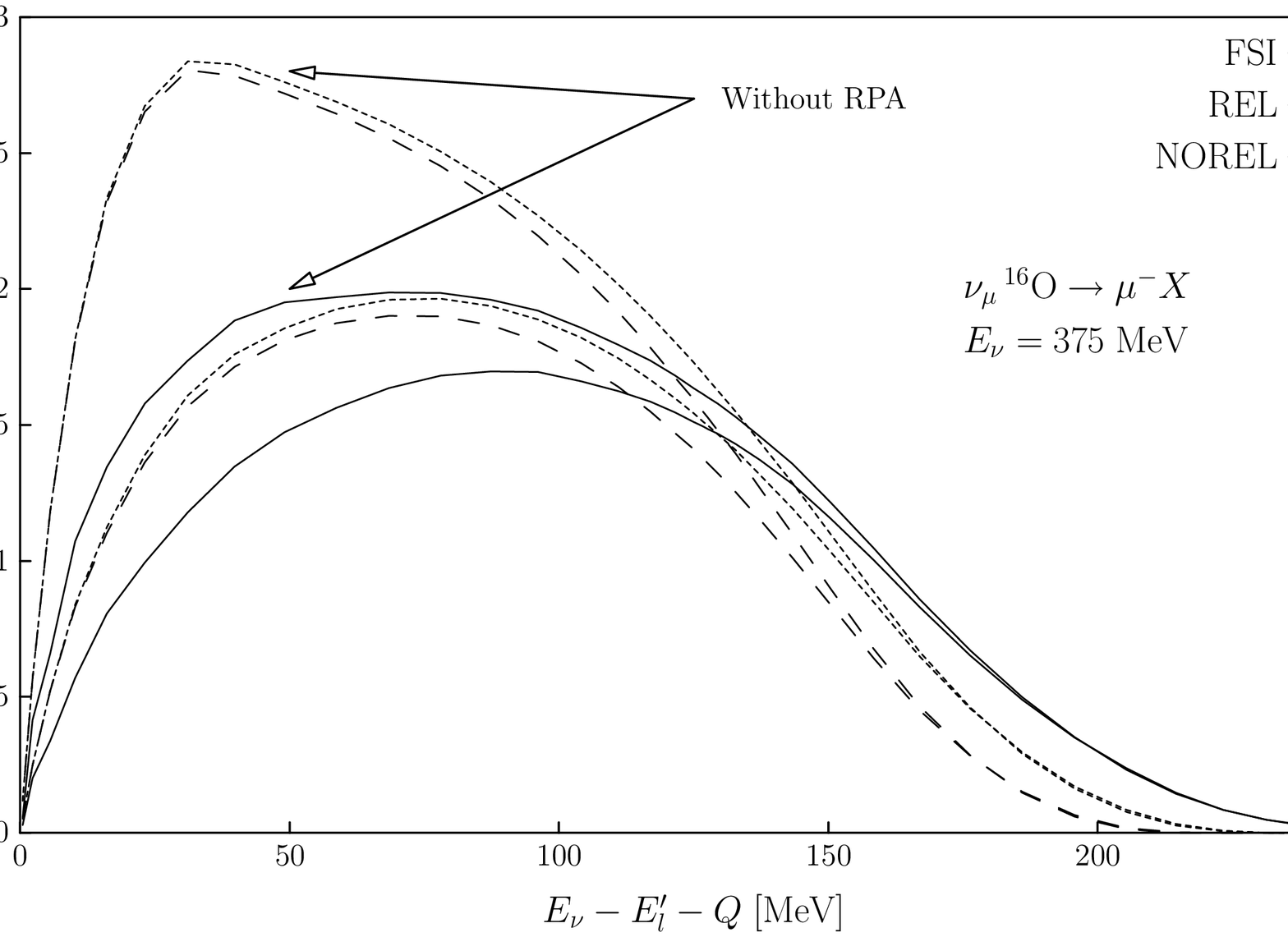}
\end{center}
\vspace{-6cm}
\caption{Muon neutrino QE differential cross
  sections in $^{16}$O as a function of the energy transfer.  
  We show results for relativistic (long dashed line, 'REL') and
  non-relativistic nucleon kinematics with (solid line, 'FSI')
  and without (short dashed line, 'NOREL') FSI effects. 
  We also show the effect of RPA and Coulomb corrections 
  (lower lines at the peak).}
\label{fig:fsi3}
\end{figure}

In Figs.\ \ref{fig:secdouble} and~\ref{fig:fsi3} FSI effects on differential
cross sections are shown. As expected, FSI provides a
broadening and a significant reduction of the strength of the QE
peak. 
Finally, in Table~\ref{tab:fsi} we compile muon and electron neutrino
and antineutrino inclusive QE integrated cross sections from
oxygen. We present results for relativistic (REL) and non-relativistic
nucleon kinematics and in this latter case, we present results with (FSI)
and without FSI (NO-REL) effects. Though FSI changes importantly the shape of
the differential cross sections, it plays a minor role when one
considers total cross sections. When medium polarization effects are
not considered, FSI provides significant reductions (13-29\%) of the
cross sections~\cite{Bl01}. However, when RPA corrections are included,
the reductions becomes more moderate, always smaller than 7\%;
even there exist some cases where FSI enhances the cross
sections. This can be easily understood by looking at
Fig.\ \ref{fig:fsi3} where we show the differential cross section as a
function of the energy transfer for $E_\nu=375$ MeV. There, we see
that FSI increases the cross section for high energy transfer. But for
nuclear excitation energies higher than those around the QE peak, the
RPA corrections are certainly less important than in the peak
region. Hence, the RPA suppression of the FSI distribution is
significantly smaller than the RPA reduction of the distribution
determined by the ordinary Lindhard function.

\section{Previous results}
  
The same formalism presented here has been used in previous works studying 
real~\cite{Nucl. Phys. A Carrasco} and virtual~\cite{Nucl. Phys. A Gil}
photon inclusive nuclear reactions.
Excellent results both in the quasielastic and $\Delta$ excitation
regions where obtained in these works.
To describe the $\Delta$ peak and the "dip" regions, they included a
high number of gauge boson absorption modes so they were able to study
the reaction at higher nuclear excitation energies than those we have
presented here. As can be seen for instance in Fig.\ \ref{fig:eegil} 
the agreement with experiment is excellent. 
Furthermore, inclusive processes of the type $(e,e^\prime N)$, 
$(e,e^\prime NN)$, $(e,e^\prime \pi)$
were studied by means of a MonteCarlo simulation as presented in
\cite{Gil-MC} that make use of the nuclear and pion physics models 
of~\cite{Phys. Rep. Oset} and~\cite{var_pion}.
\begin{figure}
\begin{center}
\includegraphics[scale=0.45]{eecarbon.ps}
\end{center}
\caption{Double differential $^{12}$C($e,e^\prime$)$X$ cross section} 
\label{fig:eegil}
\end{figure}
It is because of the remarkable succes of this model that we expect our work
to be highly reliable for CC reactions.

\section{Conclusion and Outlook}

We have presented here a many body approach to inclusive electroweak 
reactions in nuclei, at intermediate energies (nuclear excitation
energies below 500 MeV). It systematically takes into account 
RPA, SRC, $\Delta$(1232), FSI and MEC effects.
The meson-nucleon and nucleon-nucleon dynamics of the approach have been 
successfully tested in former pionic reactions.

It has been tested succesfully in:
\begin{itemize}
  \item Real and virtual photo-absorption and $\pi$, $N$, $NN$, $N\pi$
        electro and photoproduction processes in nuclei.
  \item Charged current induced inclusive neutrino 
    $^{12}$C$(\nu_\mu,\mu^-)X$ cross sections at low energies
         and Inclusive Muon capture in Nuclei.       
\end{itemize}

Predictions for QE neutrino induced reactions in nuclei at intermediate 
energies of interest for future neutrino experiments have been presented.

Our intention is to improve this approach by including contributions from 
resonance degrees of freedom and MEC in the charged current reactions. 
We also want to extend this formalism to exclusive channels in neutral
currents via a MonteCarlo simulation.

\section{List of Symbols/Nomenclature}

We have used $\hbar=c=1$ units for formulas all throughout this work, however
results in tables and figures are presented in the following units unless 
otherwise noted.

 \begin{tabbing}
$k_F^n=$ Fermi momentum for neutronss \=        \kill  
$\vec{k}=$ LAB lepton momenta, MeV \> 
$E_{\vec{p}}=$ Energy of $p$ momentum lepton, MeV    \\
$\rho=$ Nuclear matter density \> $\sigma=$ Cross section, $10^{-40}$cm${^2}$\\
$\vec{\sigma}=$ Spin Pauli matrices \> $\vec{\tau}=$ Isospin Pauli matrices  \\
$k_F^n=$ Fermi momentum for neutrons \> $k_F^p=$ Fermi momentum for protons\\
$q=$ Transfered $W$ momentum\>$\Gamma=$ Muon capture widht, $10^{-4}$s$^{-1}$\\
$\Theta(x)=$ Step function  
\end{tabbing}



\end{document}